\documentclass[conference]{IEEEtran}
\usepackage{amsmath}
\usepackage{amsthm}
\usepackage{amssymb}
\usepackage{mathrsfs}
\usepackage{url}
\usepackage{caption}
\usepackage{graphicx, subfigure}
\usepackage{algorithm}
\usepackage{algorithmic}
\usepackage{cite}

\begin{document}
%
\title{User-Centric Cooperative MEC Service Offloading}

\author{\IEEEauthorblockN{Ruoyun Chen, Hancheng Lu, Pengfei Ma\\
Department of Electrical Engineering and Information Science,\\
University of Science and Technology of China, Hefei, Anhui 230027 China\\
chenryun@mail.ustc.edu.cn, hclu@ustc.edu.cn, mpf916@mail.ustc.edu.cn}
}

\IEEEtitleabstractindextext{%
\begin{abstract}
Mobile edge computing provides users with a cloud environment close to the edge of the wireless network, supporting the computing intensive applications that have low latency requirements. The combination of offloading with the wireless communication brings new challenges. This paper investigates the service caching problem during the long-term service offloading in the user-centric wireless network. To meet the time-varying service demands of a typical user, a cooperative service caching strategy in the unit of the base station (BS) cluster is proposed. We formulate the caching problem as a time-averaged completion delay minimization problem and transform it into time-decoupled instantaneous problems with a virtual caching cost queue at first. Then we propose a distributed algorithm which is based on the consensus-sharing alternating direction method of multipliers to solve each instantaneous problem. The simulations validate that the proposed online distributed service caching algorithm can achieve the optimal time-averaged completion delay of offloading tasks with the smallest caching cost in the unit of a BS cluster.
\end{abstract}

}

\maketitle

\IEEEdisplaynontitleabstractindextext

\IEEEpeerreviewmaketitle
\section{Introduction}
With more intelligent and big-data applications developed on user terminals, the resources required by the computing intensive tasks from these applications are getting higher. For example, the popular virtual reality applications consume a lot of computing and storage resources to construct realistic images. However, the user terminals with extremely limited resources are helpless. Meanwhile, offloading these tasks to the remote cloud incurs non-negligible transmission delay through the already-congested backbone network. Hopefully, mobile edge computing (MEC) \cite{abbas2017mobile, pham2020survey} deploys small servers at the edge of the network, which provides a cloud environment in proximity of the user. Then the computation-intensive tasks can be offloaded to MEC with a low response latency.

\cite{zhu2019delay, Task2019Hu, qin2020coll, Adaptive2019Samanta} have made some contributions to the MEC task offloading. MEC-assisted vehicle platooning is considered in \cite{Task2019Hu}, where the authors minimize the average total energy consumption under the constraint of meeting the deadlines of tasks, and propose a Lyapunov optimization algorithm to solve it. Amit Samanta \emph{et al.} in \cite{Adaptive2019Samanta} try to solve the service offloading problem 
considering tasks that have different delay requirements. 
Specifically, service offloading means prefetching the dependent databases and libraries, i.e., services, to support the computing-intensive tasks at first and then performing computing. The caching process during service offloading involves the joint allocation of both computing and storage resources, which brings challenges that has not been solved perfectly.

Furthermore, the wireless network plays a significant role in many end-to-end services \cite{Lu2019QOE, Fu2020Secure, L2014Virtual, Lu2015Highly}, which can also be performance bottleneck in the offloading process. But it is ignored and not well studied in existing works. Traditionally, multiple users associate with one base station (BS) and offload their tasks to the server deployed in the associated BS. The competition for limited wireless communication resources causes a non-negligible delay compared with the processing delay when the load on servers is light. Furthermore, users located at the edge of the conventional cells experience poor signal-to-interference-plus-noise ratio (SINR). With the evolution of 5G ultra-dense network, the user-centric architecture \cite{UUCN}, \cite{UCCRAN2018Pan} which replaces the traditional network-centric architecture, enables each user to be served by a proprietary virtual network. To implement this, multiple BSs form a cluster to serve a user cooperatively \cite{zhu2018stochastic}, \cite{UCDownlink2015Su}. The data sent by the user will be jointly decoded by the BS cluster. Combined with interference cancellation technology, the uplink data rate can be increased. Consequently, the wireless communication process will not be the bottleneck of delay performance.

Nevertheless, service offloading in the user-centric network brings new challenges. In order to meet the time-varying offloading demands of the typical user in a long time span and make full use of the limited resources on the edge servers, BSs in the cluster that serves a typical user should conduct service caching cooperatively. Although there exist some studies on the cooperative service caching problem \cite{EdgeCach2018Zhang}, \cite{xu2018joint}, most of them discuss the caching strategies based on the popularity of multiple users, where the individual demands are not considered. Combining the benefits of the user-centric network for offloading process, we study the service caching problem in the user-centric wireless network to provide customized offloading for each user. In a word, this paper investigates the cooperative caching problem from a long-term perspective for a typical user. The main contributions are summarized as follows:

1) We discuss the service offloading in the user-centric network. The service caching strategy is developed by minimizing the time-averaged completion delay of the offloading tasks in a long time span, for a typical user who is served by a BS cluster cooperatively.

2) In the absence of future information, a Lyapunov optimization based online algorithm is designed, which transforms the long-term service caching problem into multiple instantaneous optimization problems. 
Further to implement the cooperative service caching in each time slot, a distributed consensus-sharing algorithm under the alternating direction method of multipliers (ADMM) framework is proposed.

3) We carry out extensive simulations to validate the optimality and the superior convergence performance of the proposed algorithm. The results demonstrate that our algorithm outperforms the instantaneous optimal caching strategy and iteration based algorithm in terms of different metrics.




The rest of this paper is organized as follows. Section II describes the system model. The problem is formulated and analyzed in section III. Section IV proposes the algorithm to solve the problem. The simulations are presented in Section V. Finally, this paper is concluded in section VI.

\section{System Model}
We consider a wireless network with $M$ BSs, denoted by set $\mathcal{M}$. Each BS is equipped with $A$ antennas and endowed with cloud resources, i.e., processing and storage capabilities. Constrained by the deployment overhead, the BS $m$ ($m\in \mathcal M$) has a CPU with maximum processing capacity $C_m$ (CPU cycles per Hz) and a storage with maximum space $S_m$.

\begin{figure}[htbp]
\centering
  \includegraphics[width=0.45\textwidth]{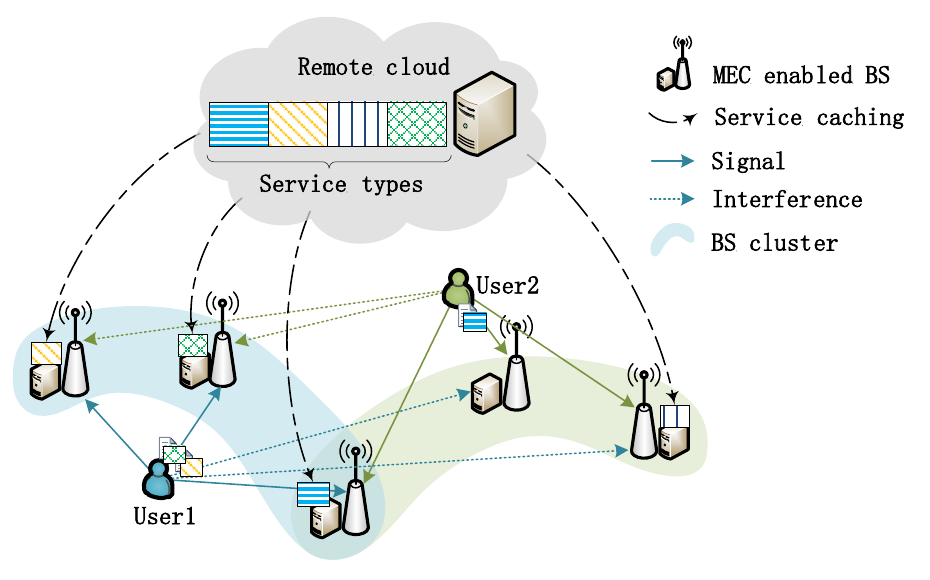}
  \caption{MEC-enabled user-centric network model.}
  \label{sysmodel}
\end{figure}

There are $U$ active users in this network, denoted by the set $\mathcal{U}$. Each user is served by a cluster of BSs. BSs in the cluster receive user data and decode it jointly by exchanging channel state information through the backhaul link. We assume that each user generates the computing tasks that request for one type of service each time. All $K$ types of the services are stored at remote cloud, denoted by the set $\mathcal{K}$. The service $k$ ($k\in \mathcal K$) has a size of $s_k$, which will take up the storage space on edge servers of size $s_k$ if it is cached. Furthermore, it has a computing requirement of $f_k$ to process per unit of data. Fig. \ref{sysmodel} describes the main elements of the system vividly. Each user is served by 3 BSs, and overlapping between users is allowed for maximizing the utility of resources.

To describe the dynamic of the communication system and users' requests, we separate the long time span $T$ into discrete time slots $t$ ($1\le t\le T,t \in \mathbb Z$), and assume that the channel state and users' requests in each time slot remain constant. The details of the system model are illustrated next.

\subsection{User-centric Communication model}
Firstly, the user-centric wireless network model requires a clear statement of BS clustering. We use a binary variable $c_{u,m}(t)$ which indicates that the BS $m$ belongs to the cluster of the user $u$ at time $t$ with value 1, otherwise 0.
The BS cluster of the user $u$ is denoted by $\Phi_u(t)$
and the users served by $\Phi_u(t)$ is denoted by $\Omega_u(t)$. Taking user $u$ as the typical user, the users belonging to $\{v:\forall v\neq u, v\in \Omega_u(t)\}$ are called intra-cluster users, while the remaining users are called inter-cluster users. With the clustering model clarified, the signal received by BS $m$ is
\begin{equation}\small
\begin{aligned}
b_m(t)=&\sum_{u\in \mathcal{U}}\sqrt{p_u}\mathbf g_{mu}(t)a_u(t)+n(t)\\
=&\sqrt{p_u}\mathbf g_{mu}(t)a_u(t)+\!\!\!\!\!\!\!\!
\sum_{\tiny{\begin{array}{c}
v\!\!\neq \!\!u,\\
v\!\!\in \!\!\Omega_u(t)\end{array}}}\!\!\!\!\!\!\!\!\sqrt{p_v}\mathbf g_{mv}(t)a_v(t)\\
&+\!\!\!\!\!\!\!\!\sum_{\tiny{\begin{array}{c}
w\!\!\neq \!\!u,\\
w\!\!\notin \!\!\Omega_u(t)\end{array}}}\!\!\!\!\!\!\!\!\sqrt{p_w}\mathbf g_{mw}(t)a_w(t)+n(t),
\end{aligned}
\end{equation}
where $p_u$ is the signal power of user $u$, $\mathbf g_{mu}(t)\in \mathcal{C}^{A\times1}$ is the complex channel coefficient between BS $m$ and user $u$. $a_u(t)$ is the complex symbol transmitted by user $u$ at time $t$ and $n(t)$ is the white Gaussian noise with variance $\sigma_n^2$ and zero mean. All the signals received by the BS $m$ can be divided into three parts: the first part is the useful signal of user $u$, the second part consists of the signals sent by intra-cluster users, i.e., intra-cluster interference, while the last part is the inter-cluster interference.

There are multiple technologies that can mitigate the interference nowadays, e.g., non-orthogonal multiple access \cite{Lu2020Distortion}. In this paper, to reflect the gain of BS clustering under the user-centric mode, beamforming (or precoding) is an effective interference cancellation technology. By designing the beamforming vector, the intra-cluster interference can be eliminated at the receiving BS cluster \cite{zhu2018stochastic}.
The projection transformation zero-forcing beamformer for user $u$ is calculated as follows
\begin{equation}\small
\begin{aligned}
\mathbf{w}_u(t)=\frac{\big(\mathbf I_{A\left|\Phi_u(t)\right|}-\mathbf G_{-u}(t)\mathbf G_{-u}^{\dagger}(t)\big)\mathbf g_u^u(t)}{\left\|\big(\mathbf I_{A\left|\Phi_u(t)\right|}-\mathbf G_{-u}(t)\mathbf G_{-u}^{\dagger}(t)\big)\mathbf g_u^u(t)\right\|_2},
\end{aligned}
\end{equation}
where $\mathbf{g}^u_v(t)=[\cdots,\mathbf g_{mv}(t),\cdots]_{m\in \Phi_u(t)}^T$ and $\mathbf{G}_{-u}(t)=[\cdots,\mathbf g_v^u(t)^T,\cdots]^T_{v\neq u,v\in \Omega_u}$.
Then we have the uplink SINR of user $u$ as
\begin{equation}\small
\begin{aligned}
S\!I\!N\!R_u\!(t)\!&=\!\frac{p_u\left|\mathbf{w}_u(t)^H\mathbf{g}^u_u(t)\right|^2}{\tiny \sum\limits_{w\notin \Omega_u(t)}
\!\!p_w\!\left|\mathbf{w}_u(t)^H\mathbf{g}^u_w(t)\right|^2\!\!+\!\left|\mathbf w_u(t)\right|^2\!\sigma_n^2}.
\end{aligned}
\end{equation}
The uplink data rate of user $u$ is $r_u(t)=log_2(1+S\!I\!N\!R_u(t))$.

\subsection{Offloading Task Model}
The task generated by user $u$ at time $t$ is denoted by $T_u(t)$, which is characterized by $(d_{T_u(t)}, w_{T_u(t)})$ representing the size of data and workload separatively. The workload refers to the computing resources required to process each unit of data, e.g., CPU cycles per bit, which can be acquired from the profile of the task. To clarify the type of service that the task requests for, we use a binary variable $o_{k,T_u(t)}$ to indicate that the task generated by user $u$ at time $t$ requests for service $k$ with value 1, otherwise 0. We assume that each user uploads a computing task that requests for one type of service at each time slot. A task which consists of multiple types of services will be considered as several independent tasks generated by different users. According to the above assumption, we have the constraint $\sum_{k\in \mathcal{K}}o_{k,T_u(t)}=1$.

\subsection{Service Offloading Model}
The process of service offloading includes uploading tasks, processing tasks, and returning the computing results. Above all, the BSs have to prefetch the services from remote cloud to support the processing according to a caching strategy at the beginning of each time slot. Then the tasks arrive in each time slot.

Let $x_{k,m}(t)\in [-1,1]$ be the caching strategy for service $k$ in BS $m$ at time $t$. It indicates caching ($x_{k,m}(t)>0$) or removing ($x_{k,m}(t)<0$) by the sign, and the absolute value represent the corresponding probability. $\mathbf x_m(t)=[x_{1,m}(t), \cdots, x_{K,m}(t)]^T$ represents the caching strategy of BS $m$. $\mathbf X^u(t)=[\mathbf x_m(t)|m\in \Phi_u(t)]$ denote the caching strategy of the BS cluster$\Phi_u(t)$.

We use $h_{k,m}(t)\in [0,1]$ to denote the probability that service $k$ has been cached on BS $m$ until time $t$. Services that cached on servers will occupy resources. Consequently, limited resources on the edge servers limit the number of services that can be cached, which is represented by the constraints $\sum_{k\in \mathcal{K}}h_{k,m}(t+1)s_k\leq S_m$ and $\sum_{k\in \mathcal{K}}h_{k,m}(t+1)f_k\leq C_m$ $\forall m \in \mathcal M$. Services can be cached or removed by a certain probability. Consequently, the next caching status is only determined by the current caching status and decision, which means that the probabilistic service caching process is a Markov decision process. The state transition function in our service caching model is
\begin{equation}\small
\begin{aligned}
\mathbb{P}[state(t+1)]&\!\!=\!\mathcal{P}(state(t),action(t))\\
&\!\!=\!\![x_{k,m}(t)]^{\!+}\!\!\!+\!h_{k,m}(t)\!-\!\left|x_{k,m}(t)\right|\!h_{k,m}(t),
\end{aligned}
\end{equation}
where the $[\cdot]^{+}$ means $max(\cdot,0)$. The last equation comes from the independence between $x_{k,m}(t)$ and $h_{k,m}(t)$.


Obtaining services from the cloud costs money, e.g., purchasing or transmitting. We assume that these costs differentiated by service types are proportional to the size of services. Let $Cost_k=\xi_ks_k$ denote the cost for prefetching service $k$, where $\xi_k$ is the cost coefficient of per unit of service data, then the expected prefetching cost of BS $m$ is calculated as
$$Cost_m(t)=\sum_{k\in \mathcal{K}}\xi_ks_k[x_{k,m}(t)]^{+}(1-h_{k,m}(t)).$$


In each time slot, the user sends the offloading task with the size of $d_{T_u(t)}$, which leads to an uplink communication delay
\begin{equation}\small
\begin{aligned}
D_{T_u(t)}^{UCN}=\frac{d_{T_u(t)}}{r_u(t)}.
\end{aligned}
\end{equation}
The data will be received by all the BSs in the cluster and decoded jointly, and the task dispatching strategy adopted in this paper is that the BS in the cluster which has the highest probability cached the requested service will process the task. Other task dispatching strategy can be integrated, but it is not our focus here. Let $\widehat k=\arg\max_ko_{k,T_u(t)}$ denote the service that $T_u(t)$ requests for, and $\widehat m=\arg\max_{m\in \Phi_u(t)}h_{\widehat k,m}(t+1)=\arg\max_mc_{u,m}(t)h_{\widehat k,m}(t+1)$ denotes the BS that will process $T_u(t)$. The processing delay on the edge server is
\begin{equation}\small
\begin{aligned}
D_{T_u(t)}^{edge}=\frac{d_{T_u(t)}w_{T_u(t)}}{f_{\widehat k}}.
\end{aligned}
\end{equation}
Meanwhile, the task will be uploaded to the remote cloud with probability $1-h_{\widehat k,\widehat m}(t+1)$. Transmitting through the backbone network incurs a large delay
\begin{equation}\small
\begin{aligned}
D_{T_u(t)}^{BKB}=\frac{d_{T_u(t)}}R,
\end{aligned}
\end{equation}
where $R$ is the data rate of the backbone network which is set as a small constant. We assume that the resources on the remote cloud is ample, thus the processing delay on the remote cloud is negligible. Then the delay that offloading to the remote cloud mainly comes from the transmission.

Generally, the results of calculations are very small, so we ignore the delay caused by returning the results. Summing up all the delay generated during the offloading process, the completion delay of $T_u(t)$ is
\begin{equation}\label{Dtotal}\small
\begin{aligned}
D_{T_u(t)}^{total}=&D_{T_u(t)}^{UCN}+\\
&h_{\widehat k,\widehat m}(t\!+\!1)D_{T_u(t)}^{edge}\!+\!\big(1\!-\!h_{\widehat k,\widehat m}(t\!+\!1)\big)D_{T_u(t)}^{BKB}.
\end{aligned}
\end{equation}

\section{Problem Formulation}
After clarifying the system model, the long-term service caching problem is formulated as a minimization of the average task completion delay of the whole offloading process for multiple time-slots in a long time span, which is illustrated by problem \textbf{P}.
\begin{subequations}\label{P1}\small
\begin{align}
\label{goal}
\mathbf{P}: &\quad \min_{\mathbf X^u(t)}\frac{1}{T}\sum_{t\in T}D_{T_u(t)}^{total},\\
\label{costC}
s.t.&\quad \frac1T\sum_{t\in T}\sum_{m\in \mathcal{M}}c_{u,m}(t)Cost_m(t)<Cost^{th},\\
\label{storeC}
&\quad \sum_{k\in \mathcal{K}}h_{k,m}(t+1)s_k\leq S_m,\\
\label{compuC}
&\quad \sum_{k\in \mathcal{K}}h_{k,m}(t+1)f_k\leq C_m,\\
\label{variC}
&\quad x_{u,m}(t)\in [-1,1].
\end{align}
\end{subequations}
Among where, $Cost^{th}$ is a hyperparameter deciding the cost threshold of the service provider, we hope for a small caching cost to guarantee the response latency during the service offloading process. Constraint (\ref{costC}) is the average service caching cost for multiple time slots. 
(\ref{storeC}) and (\ref{compuC}) are the expected storage resources and computing resources constraints on the edge servers, and (\ref{variC}) indicates the domain of the variables.

Analyzing the above problem, the service caching strategy $\mathbf X^u(t)=[\mathbf x_m(t)|m\in \Phi_u(t)]$ is the decision variable of the user-centric service caching problem. From (\ref{Dtotal}) we can tell that the delay of wireless communication is independent of the caching process, while the processing delay is affected by the caching strategy of the BS cluster, which means that the service caching strategy depends on the clustering. Consequently, we solve the service caching problem based on the optimal BS clustering \cite{clustering2014Lin}. Naturally, our optimization goal is to minimize the averaged processing delay
$$1/T\sum_{t\in T}h_{\widehat k,\widehat m}(t+1)D_{T_u(t)}^{edge}\!+\!\big(1\!-\!h_{\widehat k,\widehat m}(t+1)\big)D_{T_u(t)}^{BKB}.$$

The challenges of solving this problem come from the unknown dynamic requests of the user and the huge decision space of the BS cluster. For the unknown dynamic requests, the prediction of requests in next several time slots is generally more accurate, but it is difficult to predict the long-term requests. An online algorithm is necessary in the absence of future information. Troubled by the huge decision space, a distributed solution is more effective. Taking the above two points into consideration, we propose an online distributed service caching algorithm which is elaborated next.

\section{Algorithms: On-ConShAD}
\subsection{Online Service Caching}
Assuming that we have acquired the optimal clustering strategy for the typical user $\mathbf C^{u^{*}}(t)$ at the beginning of each time slot, then we solve the service caching problem (\ref{P1}).

The long-term cluster-based service caching problem hopes to improve the diversity of cached services through the collaborative caching between BSs, to handle the phenomenon that services cached may not be requested immediately but repeatedly in the future. To achieve this, the caching cost should be as low as possible, which is reflected in the form of constraint (\ref{costC}) where the caching variables of different time slots are coupled. 
Based on Lyapunov optimization, we transform the long-term constraint (\ref{costC}) into a queue stability problem. Specifically, we construct a virtual caching cost queue $C(t)$ with the assumption that $C(t)=0$, indicating the deviation of current caching cost, and its dynamic evolves as follows
\begin{equation}\label{costQ}\small
C(t+1)=[C(t)+a(t)-b(t)]^{+},
\end{equation}
where $a(t)\!\!=\!\!\sum_{m\!\in\! \mathcal{M}}\!c_{u,m}\!(t)\!\Big(\!\sum_{k\!\in\! \mathcal{K}}\!\xi_k\!s_k[x_{k,m}(t)]^{+}\!\big(1\!-\!h_{k,m}\!(t)\!\big)\!\!\Big)$, and $b(t)=Cost^{th}$.

The Lyapunov function is defined as $L(t)=1/2C(t)^2$ and the Lyapunov drift is $\triangle(t)=L(t+1)-L(t)$. According to the above definitions, we have
\begin{equation}\small
\begin{aligned}
\triangle(t)&=\frac12C(t+1)^2-\frac12C(t)^2\\
&\leq\frac12\big(C(t)+a(t)-b(t)\big)^2-\frac12C(t)^2\\
&=Q(t)+C(t)\big(a(t)-b(t)\big),
\end{aligned}
\end{equation}
where $Q(t)=1/2\big(a(t)-b(t)\big)^2$, which can be proved that $Q(t)\leq \overline {Q(t)}=1/2\big(\big(\sum_{m\in \mathcal{M}}c_{u,m}(t)\sum_{k\in\mathcal{K}}\xi_ks_k(1-h_{k,m}(t))\big)^2+ (Cost^{th})^2\big)$. Consequently, the long-term constrained problem (\ref{P1}) is transformed into an instantaneous problem of minimizing an upper bound on the \emph{drift-plus-penalty} expression under Lyapunov optimization framework
\begin{subequations}\label{lya}\small
\begin{align}
\label{online-tar}
&\quad \min_{\mathbf X^u(t)}\overline {Q(t)}+C(t)\big(a(t)-b(t)\big)+VD_{T_u(t)}^{pro},\\
s.t.
&\quad (\ref{storeC}), (\ref{compuC}), (\ref{variC}),
\end{align}
\end{subequations}
where $V$ is a non-negative weight that is chosen as desired to trade off between the completion delay of the users and the caching cost of the edge computing service providers.
\subsection{Distributed Parallel Caching: An ADMM Approach}
As the dynamics of BS clustering, the service caching problem of one user involves the strategies of the whole BS set, which has the size of $K \times M$.
The large-scale problem calls for a decentralized algorithm. According to \cite{boyd2011distributed}, ADMM is a widely used algorithm for solving large-scale optimization problems. Furthermore, in order to improve the diversity of cached services with a small caching cost from the perspective of BS cluster, BSs in the same cluster should cooperate with each other. To handle it, we introduce a cooperative service caching algorithm based on the consensus-sharing framework of ADMM.

To solve the online caching problem (\ref{lya}) using ADMM, we analysis whether the optimization target (\ref{online-tar}) is decomposable at first. On one hand, the drift item can be calculated independently between each BS and then get together. It can be rewritten as the sum of the decomposed items (the time index $t$ is omitted here)
\begin{equation}\small
\begin{aligned}
f(\mathrm{X})\!\!&=\!\!\overline {Q}\!+\!\!\!\!\!\sum_{m\in \mathcal{M}}\!\!\!Cc_{u,m}^{*}\Xi^T\!\!\big([\mathbf x_m]^{+}\!\!\odot\!(1\!-\!\mathbf h_m)\!\big)\!-\!CCost^{th}\\
\!&=\!\!\!\!\!\sum_{m\in \mathcal{M}}\!\!\!\!Cc_{u,m}^{*}\!\Xi^T\!\!\big([\mathbf x_m]^{\!+}\!\!\odot\!\!(1\!\!-\!\!\mathbf h_m)\!\big)\!\!+\!\!\frac1M\!(\overline {Q}\!-\!CC\!ost^{\!th}\!)\\
&=\!\!\!\sum_{m\in \mathcal{M}}\!f_m(\mathbf x_m),
\end{aligned}
\end{equation}
where $\mathrm X$ is the caching strategy of all the BSs in the system. $\mathbf \Xi_{K\times 1}=[\xi_1s_1, \cdots, \xi_Ks_K]^T$ is the service cost vector and $\mathbf h_m(t)_{K\times 1}=[h_{1,m}(t), \cdots, h_{K,m}(t)]^T$ is the caching state vector of BS $m$, $\odot$ means that the corresponding elements of two vectors or matrices are multiplied.

On the other hand, the penalty item in (\ref{online-tar}), i.e., the expected processing delay of $T_u(t)$, is determined by the probability $h_{\widehat k,\widehat m}(t+1)=\max_m\{[x_{\widehat k,m}(t)]^{+}+h_{\widehat k,m}-\left|x_{\widehat k,m}(t)\right|h_{\widehat k,m}\}$. Hence, the penalty depends on the caching result of the whole BS cluster. This dependency prevents the penalty item from being decomposed directly into multiple agents. Therefore, we design a sharing optimization goal according to this to promote the cooperation in the consensus-sharing framework.

The penalty item in (\ref{online-tar}) can be expressed as follows
\begin{equation}\small
\begin{aligned}
P\!=D_{T_u(t)}^{pro}=\!\max_{m\in \mathcal{M}}\!\!\Big\{\!c_{u,m}\!(t)^{\!*}\mathbf{o}^{\!T}\mathbf h_m\!(t\!+\!1)\!\Big\}
(\!D_{T_u(t)}^{edge}\!-\!D_{T_u(t)}^{BKB}\!),
\end{aligned}
\end{equation}
where $\mathbf{o}_{K\!\times\!1}\!=\![o_{1,T_{\!u}\!(t)},\! \cdots,\! o_{\!K,T_{\!u}\!(t)}]^T$ is the service type indicating vector of task $T_u(t)$. 
According to the analysis of processing delay, we 
design the shared minimization goal in the form of a function related to the maximum of all the local caching variables, which is elaborated as follows (the time index $t$ is omitted here and later)
\begin{equation}\small
\begin{aligned}
g( \mathbf{L}(\mathrm{\mathbf{X}}))
=\max_{m\in\mathcal{M}}\{\mathcal{L}_m(\mathbf x_m)\}(D_{T_u(t)}^{edge}-D_{T_u(t)}^{BKB}),
\end{aligned}
\end{equation}
where $\mathcal{L}_m(\mathbf x_m)\!\!=\!\!c_{u,m}^{*}\mathbf{o}^T\!\big([\mathbf x_m]^{+}\!+\!\mathbf h_m\!-\!\left|\mathbf x_m\right|\!\odot \mathbf h_m)$ is a nonlinear function of the local caching strategy of BS $m$.

With the above transformations, we can reformulate (\ref{lya}) as a consensus-sharing problem by introducing a consensus variable set $\{z_m\!\!\in\!\! \mathbf{\mathcal R}\}_{m\in\mathcal M}$ with constraint $\mathcal{L}_m(\mathbf x_m)\!=\!z_m$.
\begin{subequations}\label{ADMM}\small
\begin{align}
&\quad \min_{\mathrm X}\!\sum_{m\in \mathcal{M}}f_m(\mathbf x_m)+Vg(\mathbf z),\\
s.t.
&\quad \mathcal{L}_m(\mathbf x_m)=z_m, \forall m\in \mathcal{M},\\
\label{admmConsS}
&\quad \big([\mathbf x_m]^{+}+\mathbf h_m-\left|\mathbf x_m\right|\odot \mathbf h_m\big)^T\mathbf{s}\leq S_m,\\
\label{admmConsC}
&\quad \big([\mathbf x_m]^{+}+\mathbf h_m-\left|\mathbf x_m\right|\odot \mathbf h_m\big)^T\mathbf{f}\leq C_m,
\end{align}
\end{subequations}
where $\mathbf{s}_{K\times 1}=[s_1, \cdots, s_K]^T$ is the size vector of services. The augmented Lagrange function is
\begin{equation}\small
\begin{aligned}
&L_{\rho}(\mathrm X, \mathbf z, \mathbf{y})=Vg(\mathbf z)+\sum_{m\in \mathcal{M}}\big(f_m(\mathbf x_m)\\
&+y_m(\mathcal{L}_m(\mathbf x_m)-z_m)+\frac\rho 2\left\|\mathcal{L}_m(\mathbf x_m)-z_m\right\|^2\big),
\end{aligned}
\end{equation}
with dual variables $y_m\in \mathbf{\mathcal{R}},\forall m \in \mathcal M$.

By defining the residual $r_m^\tau=\mathcal{L}_m(\mathbf x_m^\tau)-z_m^\tau$, the variables are updated iteratively by solving the following scaled optimization problems
\begin{subequations}\small
\begin{align}
\label{scadmm-x}
\quad \mathbf x_m^{\tau+1}:=&\arg\min_{\mathbf x_m}f_m(\mathbf x_m)+\frac\rho2\left\|r_m+u_m^\tau\right\|^2\\
\label{scadmm-z}
\quad \mathbf z^{\tau+1}:=&\arg\min_{\mathbf z}Vg(\mathbf z)+\!\frac\rho2\!\!\sum_{m\in \mathcal M}\!\left\|z_m\!-\!\mathcal{L}_m(\mathbf x_m^{\tau\!+\!1})\!-\!u_m^\tau\right\|^2\\
\label{scadmm-u}
\quad u_m^{\tau+1}:=&u_m^\tau+\mathcal{L}_m(\mathbf x_m^{\tau+1})-z_m^{\tau+1},
\end{align}
\end{subequations}
where $u_m\!=\!1\!/\!\rho\;y_m$. The minimization of (\ref{scadmm-x}) can be solved completely decentralized, i.e., each BS solves the local minimization problem (\ref{scadmm-x}) under the constraints (\ref{admmConsS}), (\ref{admmConsC}) in parallel.

Finally, the \textbf{On}line \textbf{Con}sensus and \textbf{Sh}aring \textbf{A}DMM based \textbf{D}istributed algorithm (On-ConShAD) which solves the problem (\ref{P1}) is performed as Algorithm. \ref{cachAlg}.

\begin{algorithm}[htbp]
\small
\caption{On-ConShAD.}
\label{cachAlg}
\begin{algorithmic}[1]
\STATE {\textbf{Initialization:} $C(0)=0$, $h_{k,m}(0)=0,x_{k,m}(0)=0, \forall k\in\mathcal{K},m\in\mathcal{M}$;}
\FOR {$t=1$; $t\leq T$; $t++$ }
    \STATE Acquire the optimal BS clustering $\mathbf C^{u^*}(t)$, and calculate the uplink delay;
    \STATE Observe the caching state $\mathbf h_m(t),\forall m \in \mathcal M$ of last time slot, and calculate the cost queue $C(t)$ according to (\ref{costQ});
    \STATE Predict the offloading task request $T_u(t)$.
    \STATE {\textbf{Initialization:} $\tau=0$, error tolerance $\epsilon>0$, the maximum iterations $iter^{max}$, the dual variable $\mathbf u=\mathbf 0$;}
        \REPEAT
        \STATE {\textbf{Step1}: $\forall m\in \mathcal{M}$, acquire $\mathbf x_m(t)^{\tau+1}$ by solving (\ref{scadmm-x}) under (\ref{admmConsS}), (\ref{admmConsC}) in parallel;}
        \STATE {\textbf{Step2}: Gather $\mathbf x_m(t)^{\tau+1}$ from all BSs in cluster and then update $z_m(t)^{\tau+1}$ by solving (\ref{scadmm-z}) and calculate dual variable $\mathbf u$ according to (\ref{scadmm-u});}
        \STATE {\textbf{Step3}: Set $\tau=\tau+1$;}
        \UNTIL {the termination criterion is satisfied, i.e.,\\ $\sum_{m\in \mathcal M}\left\|\mathcal{L}_m(\mathbf x_m^{\tau})-z_m(t)^\tau\right\|^2\leq\epsilon$} or $\tau>iter^{max}$
    \STATE Calculate the time-averaged caching cost and the time-averaged task completion delay.
\ENDFOR
\end{algorithmic}
\end{algorithm}

\section{Performance Evaluation}
\begin{figure}[b]
\centering
\subfigure[Time-averaged performance.]{
\begin{minipage}[t]{0.48\linewidth}
\centering
\includegraphics[width=1\textwidth]{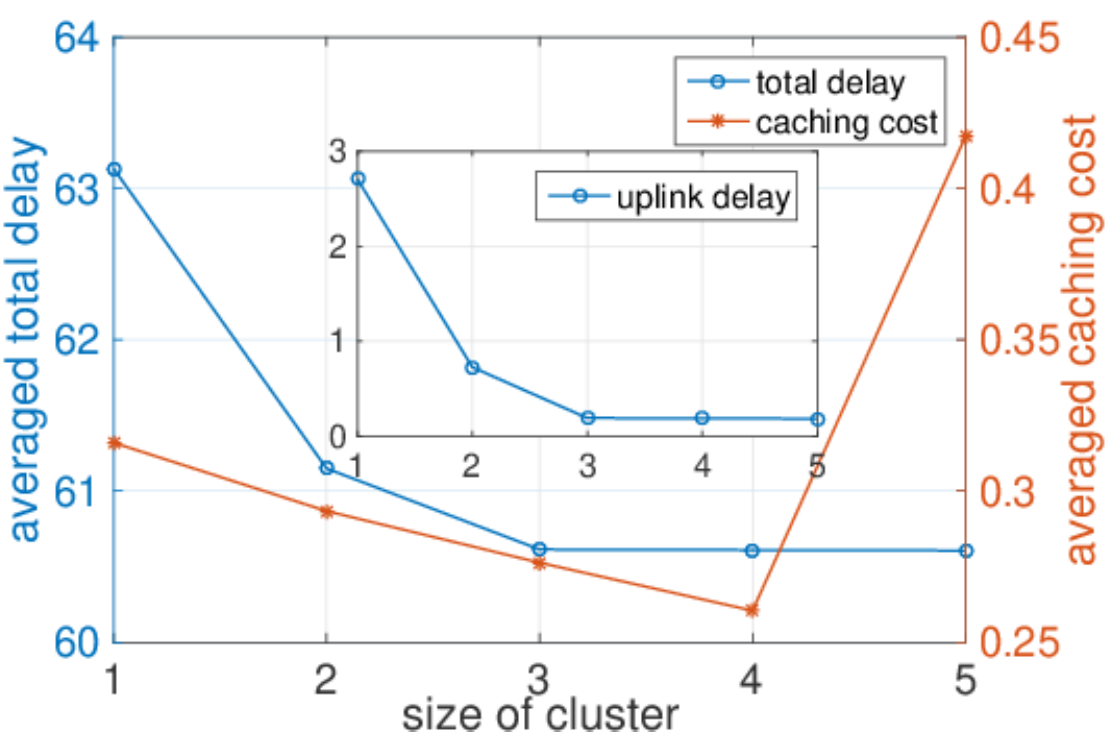}
\label{diffclus}
\end{minipage}
}
\subfigure[Convergence of ADMM.]{
\begin{minipage}[t]{0.43\linewidth}
\centering
\includegraphics[width=1\textwidth]{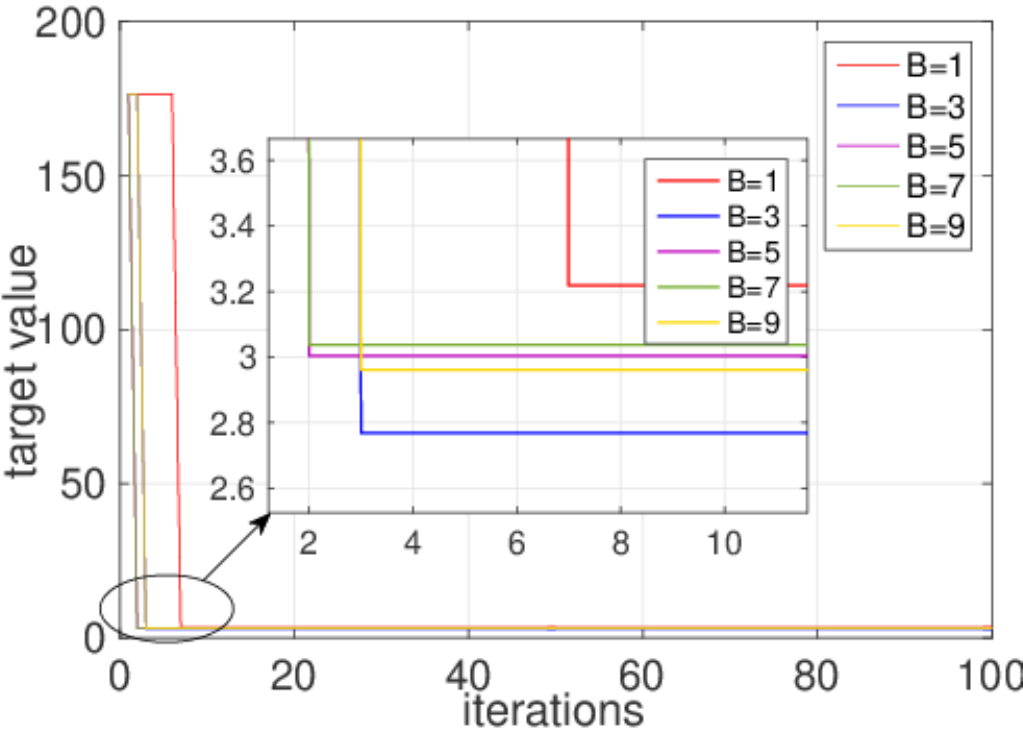}
\label{admmconver}
\end{minipage}
}
\centering
\caption{Performance under different size of BS cluster.}
\end{figure}

In this section, we evaluate the performance of On-ConShAD by some simulations. We set 10 BSs and 6 types of services. Each of the BS is equipped with 2 antennas and is endowed with a resources-limited server. Each service has specific configurations including size, cache overhead factor and resources requirement. We introduce interferences by setting 4 communication users.

First, we validate the superiority of serving user with BS cluster. 
In Fig. \ref{diffclus}, the size 1 represents the case of serving user with single BS. On one hand, the decreasing circle line shows that serving user with BS cluster has a better delay performance. It is worth noting that the trend of the total delay and the uplink delay is basically the same, which means that the contribution of the BS cluster is mainly reflected in the uplink communication process. The joint decoding and interference elimination in the clustering mode lead to a higher uplink data rate, thus the uplink communication delay is smaller. 
On the other hand, the caching cost of BS cluster keeps decreasing until the size gets up to 4 while it increases when the size is larger. This reveals that the BSs in a larger cluster are more likely to perform redundant caching because of available resources, which aims at ensuring a minimum processing delay.

Further to explore the influence of cluster's size on our algorithm, we test the convergence performance of distributed consensus-sharing ADMM. As shown in Fig. \ref{admmconver}, the convergence always takes only several iterations as the size changes, which reveals that the size of the BS cluster has slight effect on the convergence performance of the distributed algorithm, indicating that the algorithm is scalable and can be applied to large-scale problems.

\begin{figure}[b]
\centering
\subfigure[Time-averaged processing delay.]{
\begin{minipage}[t]{0.45\linewidth}
\centering
\includegraphics[width=1\textwidth]{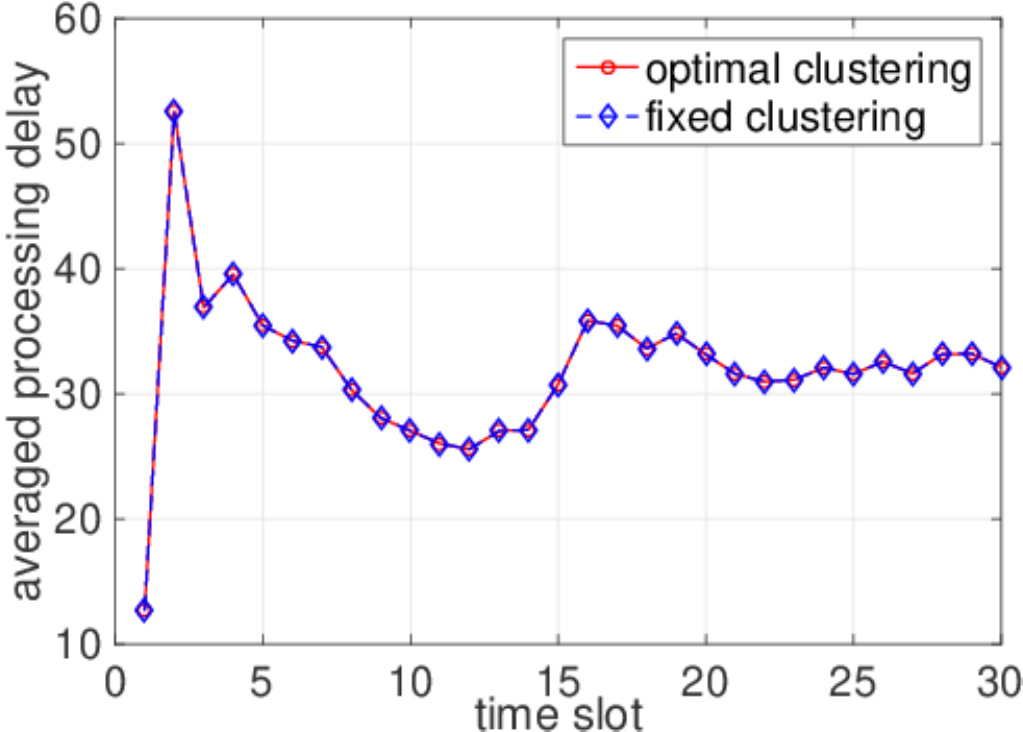}
\label{fixdelay}
\end{minipage}
}
\centering
\subfigure[Time-averaged caching cost.]{
\begin{minipage}[t]{0.45\linewidth}
\centering
\includegraphics[width=1\textwidth]{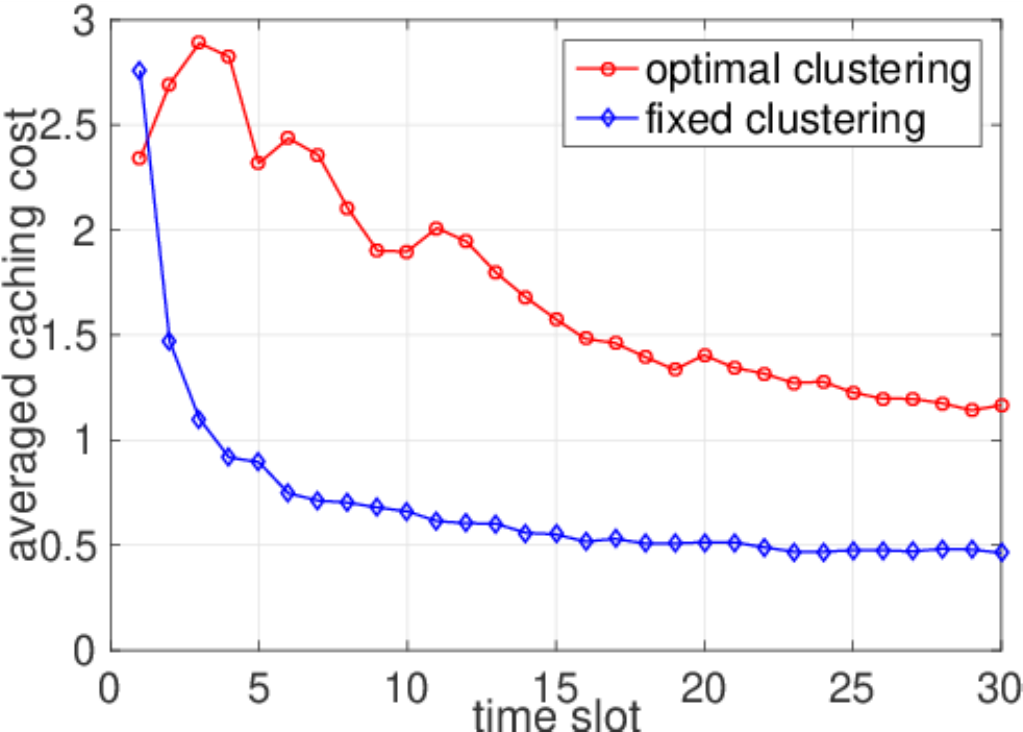}
\label{fixcost}
\end{minipage}
}
\centering
\caption{Impact of the way the BS cluster is divided.}
\label{fixtest}
\end{figure}
\begin{figure}[b]
\centering
\subfigure[Time-averaged total delay.]{
\begin{minipage}[t]{0.465\linewidth}
\centering
\includegraphics[width=1\textwidth]{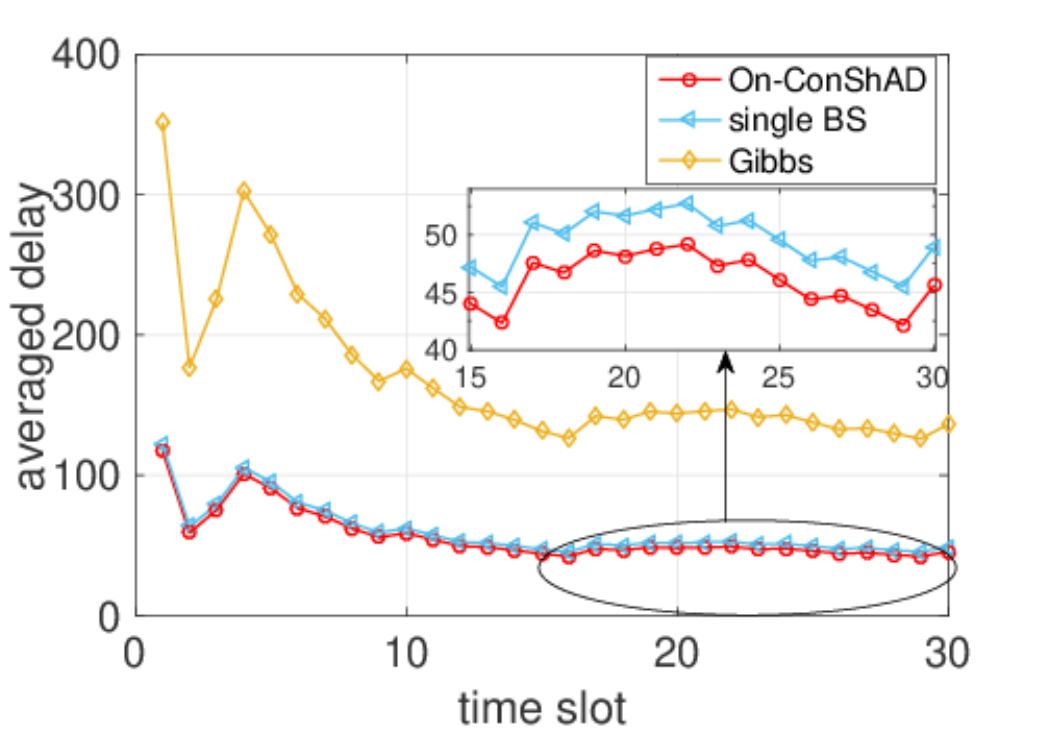}
\label{compdelay}
\end{minipage}
}
\centering
\subfigure[Time-averaged caching cost.]{
\begin{minipage}[t]{0.465\linewidth}
\centering
\includegraphics[width=1\textwidth]{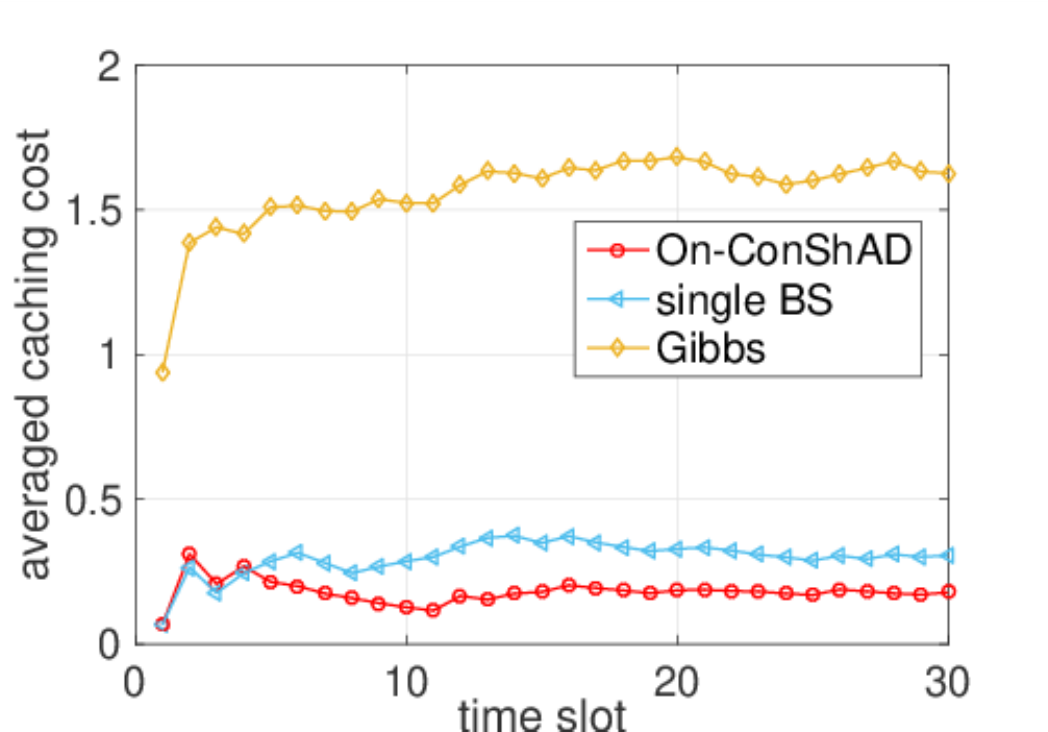}
\label{compcost}
\end{minipage}
}
\centering
\caption{Time-averaged performance comparison.}
\end{figure}
Second, although we take the advantage of the cooperations of BS clusters, we did not discuss the division of BS clusters. The proposed algorithm is based on the assumption that the BS cluster is divided optimally and dynamically. Therefore, we have to figure the degree how much the clusters' division impacts on the caching performance. In Fig. \ref{fixtest}, the circle line represents the dynamic division while the diamond line represents the fixed division. The size of cluster is set as 3. The delay performance is unaffected (Fig. \ref{fixdelay}). However, the caching performance degraded when we divide the cluster dynamically (Fig. \ref{fixcost}). Though the dynamic clustering achieves optimal uplink performance, it interrupts the long-term nature of the caching status. The services cached in the history are no longer useful due to re-clustering, which results in redundant caching. 

Further to verify the effectiveness and optimality of the proposed On-ConShAD, we set the size of BS cluster as 3 and compare it with two algorithms. \textbf{Single BS:} This is the baseline of our algorithm, in which the typical user is served by a single BS. At each time slot, the typical user chooses one BS following the rule of achieving the highest uplink data rate, and offloads to that BS. By minimizing the weighted sum of the completion delay and the caching cost on the single BS to meet user's requirement in each time slot, the solution is instantaneous optimal. \textbf{Gibbs:} This is also an iteration-based algorithm which adopts the variation of Gibbs Sampling method \cite{xu2018joint}. It makes binary caching decisions (cache with probability 1 or 0) and fractional offloading decisions (percentage of tasks computed at the edge).

Fig. \ref{compdelay} shows the time-averaged completion delay of these three algorithms. From which we can tell that the On-ConShAD performs as excellent as the instantaneous delay-optimal single BS performs. The completion delay of the tasks obtained by On-ConShAD is a little bit smaller, which has been validated in Fig. \ref{diffclus}. Fig. \ref{compcost} illustrates the effectiveness of the online strategy, where the minimization of the drift of the virtual caching cost queue limits the long-term caching cost to a small value. At first, On-ConShAD and single BS have equivalent caching cost. As time grows, single BS performs frequent replacement of services to minimize the completion delay, while On-ConShAD is gradually getting lower by minimizing the cumulative drift of the virtual caching cost queue. The poor performance of Gibbs is due to the certain caching strategy and the slow convergence of gibbs sampling. The convergence performance in \cite{xu2018joint} shows that it takes hundreds of iterations to get convergence while ADMM only takes a few iterations (\ref{admmconver}).

\section{Conclusion}
In this paper, we discussed the service offloading problem in the user-centric wireless network, where each user is served by several BSs cooperatively. We mainly studied the cooperative service caching strategy during the offloading process from a long-term perspective. An online distributed algorithm is proposed to solve the problem. At last, the simulations show the benefit of BS cluster and validate the effectiveness of the proposed algorithm. Nevertheless, there are still some limitations that can be completed in the future, e.g., the BS clustering problem should be optimized together with the caching problem thus acquiring the joint optimal offloading solutions. 

\section*{Acknowledgment}
This work was supported in part by the National Science Foundation of China (NSFC) (Grants 61771445, 61631017 and U19B2044).

\bibliographystyle{IEEEtran}
\bibliography{ref}

\end{document}